\documentclass[prd,tightenlines,floatfix,showpacs,preprintnumbers,nofootinbib,eqsecnum,superscriptaddress,twocolumn]{revtex4}
\usepackage{graphicx}  
\usepackage{dcolumn}   
\usepackage{bm}        
\usepackage{amsmath}
\usepackage{amssymb}   
\usepackage{enumitem}
\usepackage{graphicx,mathrsfs}
\usepackage{nicefrac}
\usepackage{multirow}
\usepackage{color}

\newcommand{\be}{\begin{equation}}  
\newcommand{\ee}{\end{equation}}  
\newcommand{\bea}{\begin{eqnarray}}  
\newcommand{\eea}{\end{eqnarray}}  

\newcommand{\tr}{\operatorname{tr}}

\begin{document}

\title{A new flavour imprint of $SU(5)$-like Grand Unification and its LHC signatures}
  
\author{S.~Fichet}
\email{sylvain.fichet@lpsc.in2p3.fr}
\affiliation{International Institute of Physics, UFRN, Av.\ Odilon Gomes de Lima, 1722 - Capim Macio - 59078-400 - Natal-RN, Brazil}

\author{B.~Herrmann
}
\email{herrmann@lapth.cnrs.fr}
\affiliation{LAPTh, Universit\'e de Savoie, CNRS, 9 Chemin de Bellevue, BP 110, F-74941 Annecy-le-Vieux, France}

\author{Y.~Stoll
}
\email{stoll@lapth.cnrs.fr}
\affiliation{LAPTh, Universit\'e de Savoie, CNRS, 9 Chemin de Bellevue, BP 110, F-74941 Annecy-le-Vieux, France}
                            
\date{\today}

\begin{abstract}

We point out that the hypothesis of a $SU(5)$-like supersymmetric Grand Unified Theory (GUT) implies a generic relation within the flavour structure of up-type squarks. Contrary to other well-known $SU(5)$ relations between the down-quark and charged lepton sectors, this relation remains exact in the presence of any corrections and extra operators. Moreover it remains valid to a good precision at the electroweak scale, and opens thus new possibilities for testing $SU(5)$-like GUTs. We derive the low-energy effective theory of observable light up-type squarks, that also constitutes a useful tool for squark phenomenology. We use this effective theory to determine how to test $SU(5)$ relations at the LHC. Focussing on scenarios with light stops, compatible with Natural SUSY, it appears that simple tests involving ratios of event rates are sufficient to test the hypothesis of a $SU(5)$-like GUT theory. The techniques of charm-tagging and top-polarimetry are a crucial ingredient of these tests.
\end{abstract}

\maketitle

\section{Introduction}

One of the most fascinating features of the Standard Model of Particles (SM) is that the matter fields fit into complete representations $10$ and  $\bar 5$ of the $SU(5)$ gauge group, as $10 = (Q,U,E)$, $\bar 5 = (L,D)$~\cite{Georgi:1974sy}. This suggests that the SM is the low-energy effective theory (EFT) of a $SU(5)$-like Grand Unified Theory (GUT) -- either $SU(5)$-symmetric or containing $SU(5)$ as a subgroup (see \cite{SU5GUTS1} and the reviews \cite{SU5GUTS2}). Many classes of $SU(5)$-like GUTs exist, with a variety of low-energy features.\footnote{Note that some early models are excluded by proton decay~\cite{SU5death}.} From the viewpoint of testing whether or not Nature is microscopically $SU(5)$-symmetric, this model-dependence is an irreducible theoretical uncertainty. It is thus highly challenging to find how to test the $SU(5)$ hypothesis in a way as model-independently as possible. 

In this Letter, assuming a $SU(5)$-like supersymmetric GUT, we point out a so-far unexplored $SU(5)$ property confined to  the up-squark sector, that is much less sensitive to model-dependence than previous, well-studied properties. This property arises in the flavour structure of up squarks, and its  persistence  at the weak scale is closely checked. We then derive the up-squark effective theory, applicable whenever unobserved squarks are assumed to be heavy. We use it to determine the most direct way to test  the $SU(5)$ hypothesis using our new relations, focussing on the light-stops scenario -- that is favoured by LHC and consistent with Natural SUSY.

\section{A new relation in the up-squark sector } 
 
Besides gauge-coupling unification at the GUT scale, matter field unification implies the famous relation \be y_d = y_\ell^t \ee between down quark and lepton Yukawa couplings. This  relation is exact up to  GUT scale threshold corrections that arise from integrating out heavy GUT states. In addition, the renormalization group (RG) flow down to the weak scale has to be taken into account, such that $y_d-y_\ell$  unification  is not straightforward and has generated a lot of litterature (see~\cite{Yuk_unif} and many subsequent works). 

Grand Unified Theories are closely linked to supersymmetry (SUSY), which strikingly favours the gauge coupling unification
in its most simple realizations (MSSM, NMSSM) as well as  in  more evolved hypothesis (e.g. extra-dimensions, extra $SU(5)$-like matter~\cite{SU5unif}).
 In SUSY $SU(5)$-like GUTs, the two Higgs supermultiplets,
  denoted $H_{1},\, H_{2}$ ($\equiv H_{d},\, H_{u}$), need to be embedded in a $5$ and a $\bar{5}$ representation respectively, denoted $\mathcal{H}_{1}$ and $\mathcal{H}_{2}$. Interactions between matter and Higgses are given by the superpotential
\begin{equation}
	W = \lambda_{1}^{ij} \mathcal{H}_{1} 10_i \bar{5}_j + \lambda_{2}^{ij} \mathcal{H}_{2} 10_i 10_j \,.
	\label{Eq:Superpotential}
\end{equation}

Below the GUT scale, assuming that unwanted Higgs triplets are heavy due to some splitting mechanism \cite{SU5GUTS2,DT_splitting}, the superpotential reads 
\begin{equation}
	W = y_u^{ij} H_2 Q_i U_j + y_d^{ij} H_1 Q_i D_j + y_{\ell}^{ij} H_1 L_i E_j \,.
\end{equation}
where  $y_d=y_\ell^t=\lambda_1$. Furthermore, SUSY needs to be broken around the TeV scale, and the Lagrangian also contains the SUSY-breaking scalar trilinear terms $a_{u,d,\ell}$,
\begin{equation}
	\mathcal{L}_{\rm soft} \supset a_{u}h_{u}\tilde{q}\tilde{u}+a_{d}h_{d}\tilde{q}\tilde{d}+a_{l}h_{d}\tilde{\ell}\tilde{e}\,,
\end{equation}
where $h_{u,d}$ are the two Higgs doublets and $\tilde{q}$, $\tilde{u}$, $\tilde{d}$, $\tilde{\ell}$, $\tilde{e}$ are the squarks and sleptons. The Lagrangian also contains five scalar mass terms denoted by $m^2_{Q,U,D,L,E}$ (see, e.g., \cite{Martin:1997ns}).

In the present Paper, we assume that the source of SUSY breaking is $SU(5)$-symmetric. The above terms then satisfy the relations
\begin{equation}
	a_{d} = a_{\ell}^{t}\,,\quad m^2_Q= m^2_U = m^2_E\,,\quad m^2_L = m^2_D
	\label{Eq:UsualRelations}
\end{equation}
at the GUT scale, which are exact up to GUT threshold corrections. 

The correlations that the $SU(5)$ relations \eqref{Eq:UsualRelations} induce between the quark and lepton sectors have been extensively studied within specific GUT scenarios \cite{SU5relations}. Such correlations, although certainly interesting in specific models, can be hardly used as a generic test of the $SU(5)$ hypothesis, as RG  corrections received by quark and leptons are fundamentally different.

We now point out the existence of relation implied by the $SU(5)$ matter embedding, which seems unnoticed so far. The $10_i 10_j$ term in the superpotential \eqref{Eq:Superpotential} is symmetric, such that only the symmetric part of $\lambda_{2}^{ij}$ is selected. This leads to a symmetric top Yukawa coupling at the GUT scale,
\begin{equation}
	y_{u} = y_{u}^{t} \,.
	\label{Eq:YukSymGUT}
\end{equation}
Moreover, in presence of SUSY, this enforces that the trilinear coupling is also symmetric,
 \begin{equation}
	a_u = a_u^t \,.
	\label{Eq:TrilinearUp}
\end{equation}
These relations are confined within the up-(s)quark flavour space, and can thus be expected to be more stable against quantum corrections than the quark-lepton relations. Moreover, these relations remain rigourously exact in presence of any GUT threshold corrections, because the up-squark self-energy gets in any case contracted with $10_i 10_j$ and is therefore always symmetric. \footnote{Note $y_{u}$, $a_u$ are symmetric, but generally not hermitian. }

For a non-SUSY theory, Eq.~\eqref{Eq:YukSymGUT} does not seem particularly exploitable. In this case the only physical parameters are mass eigenvalues and CKM angles, and these would not be enough to find out whether or not $y_u$ is symmetric. The situation becomes different once one considers broken SUSY, because more degrees of freedom of the Yukawa matrices are probed by the superpartners.

\section{The up-squark mass matrix}

Let us see how our new $SU(5)$ relations translate into observable properties. As the relations \eqref{Eq:YukSymGUT} and \eqref{Eq:TrilinearUp} hold within the up-squark sector, we need to scrutinize the up-squark mass term $\mathcal{L} \supset \tilde{u}^\dagger \mathcal{M}^2_{\tilde{u}} \tilde{u}$, where $\tilde{u} = (\tilde{u}_L,\tilde{c}_L,\tilde{t}_L,\tilde{u}_R,\tilde{c}_R,\tilde{t}_R)^t$ contains the six up-squarks states. 

In the super-CKM basis, defined such that the Yukawa matrices are diagonal, the up-squark mass matrix has the form
\begin{equation}
 	\mathcal{M}^2_{\tilde{u}} = 
	\begin{pmatrix}
		\hat{m}_Q^2 + O(v^2)\mathbf{1}_3 & \frac{v_u}{\sqrt{2}}\hat{a}_u + O(v M) \mathbf{1}_3 \\
 		\frac{v_u}{\sqrt{2}}\hat{a}_u^t + O(v M) \mathbf{1}_3 & \hat{m}_U^2 +O(v^2)\mathbf{1}_3 \\
	\end{pmatrix}\, ,
	\label{Eq:MassMatrix}
\end{equation}
where $M$ denotes the SUSY scale. In general, $\hat{a}_u = W^\dagger a_u V $, $\hat{m}^2_U = V^{\dagger}m^2_U V$, $\hat{m}^2_Q = W^{\dagger}m^2_Q W$, with $y_u = V^{\dagger} \hat{y}_u W$, $\hat{y}_u$ diagonal. The $SU(5)$ relation \eqref{Eq:YukSymGUT} implies $V_u^* = W_u$, so that the $SU(5)$ relations are satisfied in the super-CKM basis, $\hat{a}_u=\hat{a}_u^t$, $\hat{m}_Q^2\approx\hat{m}^2_U$. 
The mass matrix $\mathcal{M}^2_{\tilde{u}}$ involves only $a_u$, $m_Q^2$ and $m_U^2$, so that we do not have to consider the other SUSY-breaking terms at this stage. 
We assume that these parameters are real, as the low-energy bounds on CP phases (for example in the kaon system) are rather stringent. We also checked that, regardless of CP constraints, all our claims on testing $SU(5)$ remain valid for complex parameters. 

We see that the up-squark mass matrix possesses a peculiar structure above the GUT scale. However below the GUT scale this structure is potentially spoiled by the non-$SU(5)$  quantum corrections. It is thus necessary to evaluate how stable the relations $a_u=a_u^t$ and $m_Q^2\approx m_U^2$ remain upon the RG flow.  
This can be qualitatively verified using the two-loop RG equations of MSSM-like theories (\cite{Martin:1993zk}, see also \cite{SU5relations}).  The only sizeable discrepancy appears between the low-energy $m_{Q\,33}^2$ and $m_{U\,33}^2$. The other discrepancies are overall negligible with respect to other theoretical and experimental sources of uncertainty. 
This conclusion is confirmed by a numerical study of typical parameter sets using the public code {\tt SPheno} \cite{SPheno}. We find that the relative discrepancies are below 1\%, including the case of large $\tan\beta$. \footnote{In case of purely imaginary parameters, the relative discrepancies are instead of a few percents.}

We parametrize the complete up-squark mass matrix as 
\be
\mathcal{M}^2_{\tilde{u}}=\begin{pmatrix}
m^2_{11} &m^2_{12} &m^2_{13} &m^2_{14} &m^2_{15} &m^2_{16} & \\
 &m^2_{22} &m^2_{23} &m^2_{24} &m^2_{25} &m^2_{26} & \\
  & &m^2_{33} &m^2_{34} &m^2_{35} &m^2_{36} & \\
    & & &m^2_{44} &m^2_{45} &m^2_{46} & \\
    & & & &m^2_{55} &m^2_{56} & \\
        & & & & &m^2_{66} & \\
\end{pmatrix}\,.
\ee

The above low-energy $SU(5)$ relations, $a_u = a_u^t$ and $m_Q^2 \approx m_U^2$, then translate into 
\begin{align}
	m^2_{15} &\approx m^2_{24}, \quad m^2_{16}\approx m^2_{34}, \quad m^2_{26}\approx m^2_{35}, 
\label{LE_relations1}		
	\\
	m^2_{12} &\approx m^2_{45},\quad m^2_{13}\approx m^2_{46}, \quad m^2_{23}\approx m^2_{56}\,, \\
	m^2_{11} &\approx m^2_{44}, \quad m^2_{22}\approx m^2_{55} \,. \label{LE_relations3}
\end{align}

\section{Up-squark effective theory}

Although the pattern of squark masses is arbitrary in full generality, a likely situation is that the masses exhibit some hierarchy. This is favoured from naturalness considerations, from LHC bounds, as well as from certain classes of models. In such a situation the physics of the light squarks can be conveniently captured into a low-energy effective theory, where heavy squarks are integrated out.
 
Let us reorganize the up-squark mass term such that  
\begin{equation}
	\mathcal{L} \supset \tilde{u}^\dagger \mathcal{M}^2_{\tilde{u}} \tilde{u}\equiv \Phi^\dagger  \mathcal{M}^2 \Phi=
	\left(\hat{\phi}^\dagger,\phi^\dagger \right)
	\begin{pmatrix}
		\hat{M}^2 & \tilde{M}^2 \\ 
		\tilde{M}^{2\dagger} &  M^2 
	\end{pmatrix}
	\begin{pmatrix}
	\hat{\phi}  \\ 
	\phi 
	\end{pmatrix} \,,
\end{equation} 
where $\hat{\phi}$ contains the heavy states and $\phi$ the light ones. The relevant piece of the corresponding Lagrangian has the general form
\begin{equation}
	\mathcal{L}\supset \big| D\Phi \big|^2 - \Phi^\dagger \mathcal{M}^2 \Phi + \Big( \mathcal{O} \phi +\hat{\mathcal{O}} \hat{\phi} + {\rm h.c.} \Big) \,,
\end{equation}
where $\mathcal{O}$ and $\hat{\mathcal{O}}$ represent the interactions with other fields, that are potentially exploited to probe the up-squark sector.

Assuming that the eigenvalues of $\hat{M}^2$ are large with respect to the energy at which one probes the theory, the heavy squarks $\hat{\phi}$ can be integrated out, \footnote{We remind that it is not necessary to use eigenstates to perform this operation.} leaving the low-energy Lagrangian of light squarks,
\begin{equation}
	\begin{split}
		&\mathcal{L}_{\rm eff} = \big| D\phi \big|^2 \\
		& + \!\left( \! \mathcal{O} \! - \! \hat{\mathcal{O}} \big( \hat{M}^{-2} \! - \! \hat{M}^{-4} D^2 \big) \tilde{M}^2 
		\! - \! \frac{\mathcal{O}}{2} \tilde{M}^{2\dagger}  \hat{M}^{-4} \tilde{M}^2 \! \right)\! \phi + {\rm h.c.}  \\  
		& - \!\phi^\dagger \! \left( M^2  
		- \tilde{M}^{2\dagger} \hat{M}^{-2} \tilde{M}^{2}
		-\frac{1}{2}  \left\{ \tilde{M}^{2\dagger}  \hat{M}^{-4}  \tilde{M}^2, M^2 \right\} \right) \! \phi \,.  
	\end{split} \label{eq:Leff}	
\end{equation}    
In this effective Lagrangian we keep only the leading and the subleading terms of the $E^2 \hat{M}^{-2}$ expansion relevant for our purposes. Here, $E$ denotes the energy scale. It contains in principle higher dimensional couplings and derivative terms, which are either subleading or irrelevant for the observables we are going to consider, and are thus neglected. To obtain Eq.\ \eqref{eq:Leff}, one has to use the field redefinition $\Phi\rightarrow (\mathbf{1}-\frac{1}{2} \tilde{M}^{2\dagger} \hat{M}^{-4} \tilde{M}^2) \Phi$ in order to canonically normalize the light squarks. The $\{\,,\}$ is the anti-commutator. 

The imprint of the heavy up-squarks in the light up-squarks Lagrangian \eqref{eq:Leff} appears as corrections to the light up-squark masses and couplings. Physically, these corrections have to be understood both as tree-level exchange of heavy up-squarks, and as the first terms of the expansion with respect to the small parameters that describe mixing of heavy and light squarks. We emphasize that, although this effective theory approach might bear some resemblance with the ``mass insertion approximation'' (MIA) \cite{MIA}, the two approaches are fundamentally different. 
The MIA is an expansion in the limit of small off-diagonal elements of the mass matrix $\mathcal{M}^2$, i.e.\ in terms of small parameters $\mathcal{M}^2_{i\neq j}/\tr\{\mathcal{M}^2\}$. In contrast, the expansion parameter of the effective theory is $E^2 \hat{M}^{-2}$, and $\mathcal{M}^2$ can have arbitrary off-diagonal entries.

From Eq.\ \eqref{eq:Leff} we see that flavour-violating couplings of the light squarks enter at first order and are controled by $\hat{M}^{-2} \tilde{M}^2$. The flavour respecting couplings will be instead modified at the second order.
The light mass matrix $M^2$ receives a correction independent of $M^2$  at first order, and corrections proportional to $M^2$ at second order.

\section{Testing $SU(5)$-like GUTs at the LHC: the example of two light-stops}

Armed with the up-squark effective Lagrangian, we are now ready to test the low-energy $SU(5)$ relations $a_u\approx a_u^t$ and $m_Q^2\approx m_U^2$. 
In this Letter we focus on LHC physics, in the assumption that the lightest squarks are produced at the LHC, while the unobserved squarks are heavy such that the truncation of the effective Lagrangian \eqref{eq:Leff} is valid. Note that the expansion parameter behaves as the mass squared, such that for masses of, e.g., $m_\phi \sim E \sim 1~{\rm TeV}$ and $m_{\hat{\phi}} \sim 3~{\rm TeV}$, higher terms of the expansion are already suppressed by a factor of order 10. 

Supersymmetric scenarios with two light squarks that are \textit{mainly} stops, $\phi=(\tilde{t}_L,\tilde{t}_R)$, $\hat{\phi}=(\tilde{u}_L,\tilde{c}_L,\tilde{u}_R,\tilde{c}_R)$ are favoured by LHC data and are one of the features of the Natural SUSY framework \cite{Natural_SUSY}.  For simplicity, we will in the following denote these two light squarks as ``stops''. We consider $R$-parity conserving scenarios. We assume that both stops are copiously produced through flavor-diagonal processes. Indeed, gluon-initiated production of squark pairs will be the dominant production channel for squarks having a mass of about 1 TeV at the LHC with a center-of-momentum energy of 13 or 14 TeV. 

The effective Lagrangian of the stops is obtained by expanding Eq.\ \eqref{eq:Leff}, where we introduce the parameters $m_{11,44}^2\equiv \Lambda_1^2$, $m_{22,55}^2\equiv \Lambda_2^2$. We observe that the stop mass matrix depends at the leading order on $m^2_{33,66,36}$. There is no relevant information to test the $SU(5)$ hypothesis in this matrix, and the higher corrections to the mass matrix cannot be exploited either. 

The stop mass eigenstates are $(\tilde{t}_1,\tilde{t}_2)^t=R(\tilde{\theta})^t (\tilde{t}_L,\tilde{t}_R)^t$, where $R(\tilde{\theta})$ is a $SO(2)$ rotation with the stop mixing angle $\tilde{\theta}$. 
The stop mixing angle can be large and is a crucial feature of low-energy SUSY. \footnote{In particular to obtain a $125$ GeV Higgs.} Knowing its value will not be necessary in the following $SU(5)$ tests, although it will appear in intermediate steps.

Phenomenologically viable parameter configurations including sizeable symmetric off-diagonal entries in the submatrix $\hat{a}_u$, which are in agreement with constraints from flavour-changing neutral currents such as rare decays ($b\to s\gamma$, $B_s \to \mu\mu$) and $B$-meson oscillation ($\Delta M_{B_s}$) can be found in the MSSM parameter space \cite{QFV}. This statement also holds when adding the $SU(5)$ hypothesis.

\subsection{The case $m_{\tilde{t}_{1,2}} > m_{\tilde{W}} > m_{\tilde{B}}$}

As a first typical example, we consider the case where stops can decay both to the lightest neutralino, which is mostly bino-like, $\tilde{\chi}^0_1 \approx \tilde{B}$, and to a mostly wino-like second-lightest neutralino $\tilde{\chi}^0_2 \approx \tilde{W}$. This mass hierarchy is inspired by the fact that in GUTs we have the approximate relation $M_{\tilde{B}} \approx M_{\tilde{W}}/2$ at the weak scale.

Let us focus on the flavour-violating couplings of the stops, that appear at first order in Eq.\ \eqref{eq:Leff}. 
The operators that couple the stops to $\tilde{B}$ and $\tilde{W}$  in Eq.\ \eqref{eq:Leff} are $\hat{\mathcal{O}} \propto (u_L,c_L,-4\,u_R,-4\,c_R) \, \tilde{B}$ and $\hat{\mathcal{O}} \propto (u_L,c_L) \, \tilde{W}$ \cite{MSSM}. 
At first order in the effective Lagrangian (\ref{eq:Leff}), the flavour-violating couplings of the stops are proportional to
\be
	\tilde{B}\!
	\begin{pmatrix}
		\frac{m^2_{13}}{\Lambda^2_1} u_L+\frac{m^2_{23}}{\Lambda^2_2}   c_L-4\frac{m^2_{34}}{\Lambda^2_1}   			u_R-4\frac{m^2_{35}}{\Lambda^2_2}  c_R \\[2mm]
		 \frac{m^2_{16}}{\Lambda^2_1} u_L+\frac{m^2_{26}}{\Lambda^2_2}   c_L-4\frac{m^2_{46}}{\Lambda^2_1}   		u_R-4\frac{m^2_{56}}{\Lambda^2_2}  c_R 
	\end{pmatrix}\!
	R(\tilde{\theta})\!
	\begin{pmatrix}
		\tilde{t}_1 \\
		\tilde{t}_2
	\end{pmatrix},
\ee
\be
	\tilde{W}\!
	\begin{pmatrix}
		\frac{m^2_{13}}{\Lambda^2_1} u_L+\frac{m^2_{23}}{\Lambda^2_2}   c_L \\[2mm]
		\frac{m^2_{16}}{\Lambda^2_1} u_L+\frac{m^2_{26}}{\Lambda^2_2}   c_L \\
	\end{pmatrix}\!
	R(\tilde{\theta})\!
	\begin{pmatrix}
		\tilde{t}_1 \\
		\tilde{t}_2
	\end{pmatrix}.
\ee

The low-energy $SU(5)$ relations Eq.\ \eqref{LE_relations1}--\eqref{LE_relations3} directly relate these effective couplings. A numerical  analysis shows that non-$SU(5)$ discrepancies induced by the RG flow are typically of order $1\%$. The stop decay chains of interest are $\tilde{t}_{1,2}\rightarrow q\, \tilde{B} $ and $\tilde{t}_{1,2} \rightarrow q\, \tilde{W} \rightarrow q\, Z/h\, \tilde{B}$, where $q=u,c$ observed as hard jets. These two decays, occurring respectively through $U(1)_Y$ and $SU(2)_L$ interactions, are  observed separately as the final states are different. In the complete production-decay process, we ask for a flavour violating decay for one out of the two produced stops.

Let us first assume that one simply counts the amount of flavour-violating events occuring in the decays to $\tilde{B}$ and $\tilde{W}$, without disentangling between the nature of the jets nor between the original stops. It turns out that,  whenever the $SU(5)$ hypothesis is verified, and for  arbitrary stop mixing angle, both decay rates $N_Y$, $N_L$ are controlled by the same combination of parameters, according to
\be
\begin{split}
	N_{L,Y} ~\propto~ \Big( \sigma_{\tilde{t}_1} c_{\tilde{\theta}}^2+\sigma_{\tilde{t}_2} s_{\tilde{\theta}}^2  \Big) 
\Big( m_{13}^4\Lambda_1^{-4}+m_{23}^4\Lambda_2^{-4} \Big) \\+
\Big(\sigma_{\tilde{t}_1} s_{\tilde{\theta}}^2+\sigma_{\tilde{t}_2} c_{\tilde{\theta}}^2 \Big) 
\Big( m_{16}^4\Lambda_1^{-4}+m_{26}^4\Lambda_2^{-4}\Big) \\
 +2c_{\tilde{\theta}} s_{\tilde{\theta}} \big(\sigma_{\tilde{t}_1}-\sigma_{\tilde{t}_2} \Big) \Big(m_{13}^2 m_{16}^2 \Lambda_1^{-4}+m_{23}^2 m_{26}^2 \Lambda_2^{-4} \Big) . 
 \label{eq:eff_coupling1}
 \end{split}
\ee
Here, $\sigma_{\tilde{t}_i}$ denotes the inclusive cross section of the flavour-conserving production process $pp \to \tilde{t}_i \tilde{t}_i^{*}$ at LHC.

It is thus, in principle, possible to test the $SU(5)$ hypothesis using these simple decay rates. However estimating precisely the overall factors relating $N_L$ and $N_Y$ to the quantity \eqref{eq:eff_coupling1} can be challenging because this requires to know the realistic cross-section including all the kinematic selections. This drawback can be avoided using charm-tagging techniques.  The use of $c$-tagging allows to identify a fraction $N_{Y,L}^c$ of the jets due to $c$-quarks, where $N_{Y,L}=N_{Y,L}^c+N_{Y,L}^{\not c}$. The remaining fraction, $N^{\not c}_{Y,L}$ includes the jets that cannot be identified as $c$-quarks, i.e.\ up-quarks and mistagged charm-jets. As a result, whenever the $SU(5)$ hypothesis is fulfilled, the four decay rates $N_{Y,L}^{\not c}$ and $N_{Y,L}^c$ satisfy the relation
\be
	\frac{N_Y^c}{N_L^c} ~=~ \frac{N_Y^{\not c}}{N_L^{\not c}}\,.
	\label{eq:RatioYL}
\ee 
Note that this test of the $SU(5)$ hypothesis does not require to know the stop-mixing angle, nor to have a precise estimation of the stop-production cross-sections. \footnote{Note that if charm tagging efficiency is different for the $U(1)_Y$ and $SU(2)_L$ processes,  efficiency factors need to be included in the relation Eq.\ (\ref{eq:RatioYL}).}

The experimental feasibility is evaluated by computing the $p$-value from a statistical test built from Eq.~\eqref{eq:RatioYL}. 
Results are expressed in terms of equivalent Gaussian significance (see e.g.  \cite{asimov}). 
 Assuming $m_{\tilde{t}_{1,2}}=700$ GeV, 300 fb$^{-1}$ of integrated luminosity, $30\%$ of charm-tagging efficiency, and flavour-violating branching ratios of $10\%$ ($3\%$), a \textit{relative} discrepancy in the above relation can be assessed up to $15\%$ ($30\%$) with $3\sigma$ significance. For $m_{\tilde{t}_{1,2}}=1000$ GeV, a discrepancies of $50\%$ can be probed with $3\sigma$ significance for $10\%$ branching ratios, while no power remains if  branching ratios are below $3\%$.

\subsection{The case $m_{\tilde{W}}>m_{\tilde{t}_{1,2}}>m_{\tilde{B}}$}

As a second example, let us assume that the stops can only decay into the lightest neutralino $\tilde{\chi}^0_1 \approx \tilde{B}$. Using only the information from flavour-violating decays into $\tilde{B}$+jets, it turns out it is not possible to test the $SU(5)$ hypothesis,  even if one disentangles the decays of $\tilde{t}_1$ and $\tilde{t}_2$ and knows the value of the stop mixing angle.
 
For this case, let us consider the flavour-conserving interactions of the stops. The relevant operator coupling to the stops in Eq.\ \eqref{eq:Leff} is $\mathcal{O}\propto(t_L,-4\,t_R)\,\tilde{B}$. 
The decays of interest for our $SU(5)$ testing purpose are thus $\tilde{t}_{1,2} \rightarrow \tilde{B}\, t_{L,R}$. For the stops decaying into top quarks, techniques of top polarimetry \cite{Top_polarization} potentially provide a way to distinguish between decays into top of left and right chirality, $t_L$ and $t_R$. Using kinematic information it is also possible to disentangle $\tilde{t}_1$ and $\tilde{t}_2$ decays. Considering that one can both distinguish between the original stops $\tilde{t}_1$, $\tilde{t}_2$ and the outgoing $t_L$, $t_R$, we end up with four observable decay rates $N_{1,L}$, $N_{1,R}$, $N_{2,L}$, and $N_{2,R}$. 
 
At leading order, the matrix coupling the stops to $\mathcal{O}$ is unitary, $\mathcal{O} R(\tilde{\theta}) (\tilde{t}_1,\tilde{t}_2)^t $. The four decay rates satisfy thus two non trivial relations, conveniently chosen as 
\be
	\frac{N_{1,L}}{N_{1,R}} = \frac{ 1}{ 16^2} \frac{N_{2,R}}{N_{2,L}} , ~~~
 	16\left(\frac{N_{1,L}}{\sigma_{\tilde{t}_1}}\!+\!\frac{N_{2,L}}{\sigma_{\tilde{t}_{2}}}\right) =
	\frac{N_{1,R}}{\sigma_{\tilde{t}_1}}\!+\!\frac{N_{2,R}}{\sigma_{\tilde{t}_2}}\,.
\ee
Note that the various ratios of decay rates provide inequivalent measurements of the stop mixing angle. A third relation coming from the overall normalization also exists but is related to the total cross-section, that needs to be estimated precisely, and which is not necessary for our purpose of testing the $SU(5)$ hypothesis. The relevant information arises instead at next-to-leading order in the distortion of the coupling matrix in the stop effective theory. Whenever the $SU(5)$ hypothesis is true, the coupling takes the form
\be
\tilde{B}
\begin{pmatrix}
t_L\,, & -4\,t_R
\end{pmatrix}
\begin{pmatrix}
1-a 
  & -b  \\ -b  & 1-a
\end{pmatrix}\,
R(\tilde{\theta})
\,\begin{pmatrix}
\tilde{t}_1\\\tilde{t}_2
\end{pmatrix} \label{eq:NLO_stop_coupling}
\ee
with
\be
a=\frac{1}{2}\left(\frac{m_{13}^4}{\Lambda_1^4}+\frac{m_{23}^4}{\Lambda_2^4}+\frac{m_{34}^4}{\Lambda_1^4}+\frac{m_{35}^4}{\Lambda_2^4}\right)\,,
\ee
\be
b=\frac{1}{2}\left(\frac{m_{13}^2m_{16}^2}{\Lambda_1^4}+\frac{m_{23}^2m_{26}^2}{\Lambda_2^4}+\frac{m_{34}^2m_{46}^2}{\Lambda_1^4}+\frac{m_{35}^2 m_{56}^2}{\Lambda_2^4}\right)\,.
\ee

The correction to the coupling matrix in Eq.\ \eqref{eq:NLO_stop_coupling} induces a slight overall decrease in the stop decay rate. One quick way to see this is that the determinant of the coupling matrix is smaller than unity. This slight global decrease is not observable as it is cancelled by the total width in the branching ratio expression.  The crucial signature lies instead in the matrix structure. If the $SU(5)$ hypothesis is true, the distortion is symmetric, which implies that 
\be
	\frac{N_{1,L}}{N_{1,R}} \neq \frac{1}{16^2} \frac{N_{2,R}}{N_{2,L}} , ~~~
	16\left( \! \frac{N_{1,L}}{\sigma_{\tilde{t}_1}} + \frac{N_{2,L}}{\sigma_{\tilde{t}_{2}}} \!\right) =
	\frac{N_{1,R}}{\sigma_{\tilde{t}_1}} + \frac{N_{2,R}}{\sigma_{\tilde{t}_2}}. \label{eq:LR_relation}
\ee
Instead, if the $SU(5)$ hypothesis is not true, \textit{both} relations are not satisfied. Again, this test uses only ratios of decay rates, and thus does not depend crucially on the overall normalization.

Building a statistical test from Eq.~\eqref{eq:LR_relation}, assuming $m_{\tilde{t}_{1,2}} = 700$ GeV ($1000$ GeV), 300 fb$^{-1}$ of integrated luminosity and maximal stop mixing, a \textit{relative} discrepancy in the first relation of Eq.\ \eqref{eq:LR_relation} can be detected up to $9\%$ ($29\%$) with a $3\sigma$ significance.

\section{Summary}

In this Letter we point out the existence of $SU(5)$-like GUT relations that have remained  unexplored so far. These new $SU(5)$ relations are insensitive to GUT threshold corrections and are confined to the flavour structure of the up-squark sector, which makes them more stable with respect to quantum corrections than the other well-known $SU(5)$ relations. 
Due to these features, the new relations open new ways to test whether Nature is microscopically $SU(5)$-symmetric. We focus on searches at the LHC using unprecedented simple tests involving the properties of up-type squarks. 

We set up the effective theory for light up-type squarks, that is also a useful tool beyond our $SU(5)$-testing purpose. Using this effective theory, we study the case of two light stops, that is motivated by the Natural SUSY framework, to set up tests of the $SU(5)$ hypothesis. We show that fairly simple statistical tests, independent of the stop mixing angle and involving only ratios of number of events, can be set up for various mass ordering scenarios. The techniques of charm-tagging and top polarimetry play a crucial role in these $SU(5)$ tests.  More evolved statistical techniques for arbitrary low-energy spectra will be presented \cite{TO_APPEAR}.


\end{document}